\documentclass[11pt]{article}
\usepackage{jcappub}
\usepackage{graphicx} 
\usepackage[T1]{fontenc} 

\usepackage{amssymb}
\usepackage{graphicx, epsfig, bm, amsmath,mathtools}

\usepackage{hyperref}
\usepackage{cleveref}

\usepackage{subcaption}
\usepackage{soul}
\usepackage{float}
\usepackage{color}

\newcommand{\bea}{\begin{eqnarray}}
\newcommand{\eea}{\end{eqnarray}}
\def\[{\left[}
\def\]{\right]}
\def\({\left(}
\def\){\right)}

\def\d{{\rm d}}

\title{Probing the local primordial non-Gaussianity with clustering of FRB dispersion measure and galaxies}
\author[a]{Indranil Das,}
\author[a]{Gilbert Holder}

\affiliation[a]{Illinois Center for Advanced Studies of the Universe \& Department of Physics, University of Illinois at Urbana-Champaign, Urbana, IL 61801, U.S.A.}

\emailAdd{idas3@illinois.edu}
\emailAdd{gholder@illinois.edu}

\abstract{We forecast the sensitivity of a joint fast radio burst (FRB) and galaxy survey to local primordial non-Gaussianity using the full $3\times2$-point statistic: the angular dispersion-measure (DM) auto-spectrum, the DM-galaxy cross-spectrum, and the galaxy auto-spectrum. Local primordial non-Gaussianity imprints a scale-dependent bias on both the free-electron field probed by FRB dispersion measures and the galaxy field. Hence, all three spectra carry information on the local non-Gaussianity amplitude $f_{NL}$. Adopting a DESI-like luminous-red-galaxy sample and an FRB population over $35\%$ of the sky, we forecast $\sigma(f_{NL}) \simeq 5$ for both $N_{\text{FRB}} = 10^{4}$ and $10^{5}$ at fixed \textit{Planck} 2018 cosmology, marginalised over three astrophysical bias parameters $(b_e^{0}, z_{\text{fb}}, b_g^{\text{amp}})$. The $f_{NL}$ constraint is robust to marginalising over the six $\Lambda$CDM parameters $(\omega_b, \omega_{\text{cdm}}, h, A_s, n_s, \tau)$ with \textit{Planck} priors. At these sample sizes, the galaxy auto-spectrum sets the constraint, while the DM sector's contribution grows with the burst count and becomes decisive for the larger samples anticipated in the SKA era, reaching $\sigma(f_{NL}) \simeq 2$ at $10^{7}$ bursts.}

\begin{document}
\maketitle
\flushbottom

\section{Introduction}\label{section introduction}
%
The statistical properties of the primordial density field are among the few direct probes of the physics of inflation. The primordial curvature perturbations in the most basic single-field slow-roll models are almost Gaussian, with any local non-Gaussianity suppressed to the levels of the slow-roll parameters \cite{Maldacena2003}. A robust detection of primordial non-Gaussianity (PNG) of the local type at the level $|f_{NL}| \gtrsim 1$ would therefore be difficult to reconcile with the simplest single-field scenarios, pointing instead to additional fields or non-canonical dynamics during inflation \cite{Bartolo2004}. The \textit{Planck} measurement of the cosmic microwave background (CMB) bispectrum \cite{Planck2018NG} provides the tightest constraint to date, $f_{NL} = -0.9 \pm 5.1$. As a two-dimensional screen, the CMB is approaching its cosmic-variance limit, so further progress must come from the three-dimensional volume mapped by large-scale structure (LSS).

The observational handle that LSS offers on local PNG is scale-dependent bias \cite{Dalal2008, MatarreseVerde2008, Slosar2008}. Local PNG induces a correction to the bias of a tracer that scales as $k^{-2}$, so it grows steeply toward the largest scales ($k \to 0$), coupling the long-wavelength gravitational potential to the small-scale fluctuations that seed collapsed structures. For a Gaussian bias $b$, this correction is proportional to $(b-1)$, and hence strongly biased tracers such as luminous red galaxies (LRGs) and quasars are especially sensitive to $f_{NL}$. Current large-scale-structure measurements have reached $\sigma(f_{NL}) \simeq 21$ from eBOSS quasars \cite{Mueller2022} and $\sigma(f_{NL}) \simeq 9$ from the DESI first data release \cite{Chaussidon2024}, while forecasts for forthcoming surveys project $\sigma(f_{NL}) \sim 1$, particularly for multi-tracer analyses \cite{Sullivan2023}. At this precision, control of observational systematics on the largest scales, where the scale-dependent bias is strongest, becomes the limiting factor, and tracers of the same density field whose large-scale systematics differ from those of galaxy surveys are of considerable value.

Fast radio bursts (FRBs) are extragalactic radio transients of millisecond duration \cite{Lorimer2007, Petroff2022} whose signals are dispersed along the line of sight by free electrons. The resulting dispersion measure (DM) integrates the free-electron density and has already been used to locate the Universe's missing baryons \cite{Macquart2020}. Its fluctuation field is a projected, line-of-sight tracer of the electron distribution, and hence of the underlying matter, analogous to weak gravitational lensing \cite{McQuinnAJL}. Crucially, the FRB field probes the same density modes through a different set of systematics \cite{Sharma2025}. In \cite{Reischke2021}, it was shown that the DM auto-spectrum alone carries the scale-dependent bias imprint of local PNG. Since then, the DM-galaxy angular cross-correlation has been detected at high significance \cite{Wang2025}, demonstrating the observational feasibility of the approach.

In this work, we forecast the sensitivity of the full FRB-galaxy $3\times2$-point statistic to local primordial non-Gaussianity. The DM auto-spectrum alone has been used to constrain $f_{NL}$ \cite{Reischke2021}, and the $3\times2$-point formalism for FRB DMs and galaxies has been developed to forecast baryonic feedback and cosmology \cite{Sharma2025}. Here, we extend this approach to local primordial non-Gaussianity by combining the DM auto-spectrum, the DM-galaxy cross-spectrum, and the galaxy auto-spectrum into a joint constraint on $f_{NL}$. We forecast the $1\sigma$ sensitivity about a Gaussian fiducial ($f_{NL} = 0$), computing Fisher constraints for each observable combination as a function of the FRB sample size, examining the robustness of the $f_{NL}$ constraint to marginalisation over astrophysical bias and cosmological parameters, and identifying the physical mechanisms that set the relative constraining power of the DM auto-correlation, the galaxy auto-correlation, and their cross. We consider $N_{\text{FRB}} = 10^{4}$ and $10^{5}$ as our reference sample sizes, which are within the reach of next-generation FRB experiments such as CHIME Outriggers, CHORD, and DSA-2000 \cite{frbcollaboration2025chimefrboutriggersdesignoverview, CHORD, hallinan2019dsa2000radiosurvey}. We extend the forecast to $N_{\text{FRB}} = 10^{6}$ and $10^{7}$ to cover the SKA era, whose projected detection rate of order $10^{6}$ bursts per year \cite{wei2024SKA} places these samples within one year to a decade of operation. We find that the full $3\times2$-point statistic yields $\sigma(f_{NL}) \simeq 5$ for both $N_{\text{FRB}} = 10^{4}$ and $10^{5}$ over $35\%$ of the sky, improving to $\sigma(f_{NL}) \simeq 2$ for $N_{\text{FRB}} = 10^{7}$.

The paper is organised as follows. In \cref{section Theoretical framework for correlation statistics} we present the formalism for the DM and galaxy auto- and cross-correlations, the associated covariances, and the Fisher forecast. In \cref{section Survey specifications and numerical setup} we specify the survey configurations and numerical implementation. We present our results in \cref{section Results and discussions} and conclude in \cref{section Conclusions}. Throughout, we adopt a \textit{Planck} 2018 fiducial cosmology \cite{Planck2018params}.

\section{Formalism for correlation statistics}\label{section Theoretical framework for correlation statistics}
%
In this section we describe the underlying theoretical framework for the auto and cross-correlation statistics of the galaxies and FRB dispersion measures. We first introduce the dispersion measure and its relation to the free-electron field (\cref{subsection Dispersion measure}), then the scale-dependent bias induced by local primordial non-Gaussianity (\cref{Primordial Non-Gaussianity and bias}). We next construct the angular auto- and cross-spectra (\cref{subsection Angular power spectrum}) and their covariances (\cref{subsection covariance matrices}), and finally describe the tomographic decomposition (\cref{subsection Tomography}) and the Fisher formalism used for the forecasts (\cref{subsection Fisher}).

%
\subsection{FRB Dispersion Measure}\label{subsection Dispersion measure}
%
As electromagnetic waves propagate through plasma, their group velocity exhibits a frequency dependency, which in turn leads to a frequency-dependent arrival time \cite{Kulkarni2020}. Hence, the time delay of the signal of an FRB at redshift $z$ in direction $\hat{n}$ travelling through the ionised intergalactic medium (IGM) can be written as
\bea \label{time delay eqn}
\delta t(\hat{n},z) \propto \text{DM}_{\text{tot}}(\hat{n},z) \nu^{-2}
\eea
where $\text{DM}_{\text{tot}}(\hat{n},z)$ is the total DM, which is the free-electron column density integrated along the line of sight, and is the quantity of actual cosmological interest.
\newline
The total dispersion measure can be split into the contributions from the electrons distributed in the large-scale structure (LSS), the Milky Way (MW) and the host galaxy
\bea \label{DM total eqn}
\text{DM}_{\text{tot}}(\hat{n},z) = \text{DM}_{\text{LSS}}(\hat{n},z) + \text{DM}_{\text{MW}}(\hat{n}) + \frac{\text{DM}_{\text{host}}(z)}{1+z}
\eea
where $\text{DM}_{\text{host}}$ is the DM associated with the host galaxy of the FRB in the host rest frame. For the line of sight not intersecting with the galactic plane, the contribution to the DM from the Milky Way can be removed with $\lesssim 10 \text{ pc cm}^{-3}$ uncertainty \cite{NE2001a,NE2001b,ProchaskaZheng2019,DSA110CGM2023}.

The dispersion measure contribution from the large-scale structure is given as an integral over the electron number density $n_e$ along the line of sight
\bea \label{DM LSS short eqn}
\text{DM}_{\text{LSS}}(\hat{n},z) = \int^{z}_{0} dz' \, \frac{n_e(\hat{n},z')}{1+z'} \frac{c}{H(z')}
\eea
where $H(z) = H_0 \sqrt{\Omega_{m0}(1+z)^{3} + \Omega_{\Lambda 0}}$ is the Hubble parameter, $H_0$ is the Hubble constant, $\Omega_{m0}$ and $\Omega_{\Lambda 0}$ are the matter and dark energy densities in the present epoch, respectively, and $c$ is the speed of light. Decomposing the electron number density into a mean and a fluctuation, and writing the mean electron number density in terms of the baryon density in the present epoch, eqn \ref{DM LSS short eqn} can be written as \cite{Sharma2025}
\bea \label{DM LSS full equation}
\text{DM}_{\text{LSS}}(\hat{n},z) = \frac{3 c H_0 \Omega_{b0}f_e}{8\pi G m_p} \int^{z}_{0} dz' \frac{f_{\text{IGM}}(z')(1+z')[1 + \delta_e(\hat{n},z')]}{\sqrt{\Omega_{m0}(1+z')^{3} + \Omega_{\Lambda 0}}}  
\eea
where $\Omega_{b0}$ is the baryon density in the present epoch, $f_e = Y_{\text{H}} + Y_{\text{He}}/2 \approx 1 - Y_{\text{He}}/2$ is the electron fraction, $Y_{\text{H}} = 0.75$ and $Y_{\text{He}} = 0.25$ are the mass fractions of hydrogen and helium respectively, $G$ is the universal gravitational constant, $m_p$ is the proton mass, and $f_{\text{IGM}}$ is the fraction of baryons in the intergalactic medium (IGM).

One can redefine eqn \ref{DM LSS full equation} and split the $\text{DM}_{\text{LSS}}(\hat{n},z)$ into a mean and fluctuation part
\bea\label{DM mean and fluctuation and weight equation}
\text{DM}_{\text{LSS}}(\hat{n},z) &=& \int^{z}_{0} W_{DM}(z')[1 + \delta_e(\hat{n},z')] dz'\\
&=& \langle\text{DM}_{\text{LSS}}\rangle(z) + \delta_{\mathcal{D}}(\hat{n},z) \nonumber
\eea
where $W_{DM}(z) = \frac{3 c H_0 \Omega_{b0}f_e f_{\text{IGM}}(z)(1+z)}{8\pi G m_p \sqrt{\Omega_{m0}(1+z)^{3} + \Omega_{\Lambda 0}}}$, $\langle\text{DM}_{\text{LSS}}\rangle(z)$ is the mean and $\delta_{\mathcal{D}}(\hat{n},z)$ is the fluctuation part respectively. Given a normalised source redshift distribution $n(z)$ and the associated distance distribution $n(\chi) = n(z)dz/d\chi$, $\delta_{\mathcal{D}}(\hat{n},z)$ can be averaged over the redshift \cite{Reischke2021}
\bea\label{DM fluctuation equation}
\delta_{\mathcal{D}}(\hat{n}) = \int^{\chi_{H}}_{0} d\chi\, W_{\mathcal{D}}(\chi) \delta_{e}(\hat{n},z(\chi))
\eea
where $\chi_{H}$ is the comoving horizon distance, and $W_{\mathcal{D}}(\chi)$ is the averaged weighting function given by
\bea\label{DM average weighting function eqn}
W_{\mathcal{D}}(\chi) = W_{DM}(\chi) \int^{\chi_{H}}_{\chi} d\chi' \, n(\chi')
\eea
We do not include source clustering, which \cite{Alonso2021, Wayland2026} show to be a negligible contribution.
%

%
\subsection{Primordial non-Gaussianity and bias}\label{Primordial Non-Gaussianity and bias}
%
The simplest models of inflation, with a single, slow-roll scalar field, predict Gaussian initial conditions. Any detection of primordial non-Gaussianity (PNG) would signal a departure from this minimal picture \cite{Maldacena2003, Bartolo2004}. We consider local transformations of the underlying Gaussian field $\phi_G$ and Taylor-expand the transformation \cite{KomatsuSpergel}
\bea\label{non gaussianity field equation}
\phi_{NG} = \phi_{G} + f_{NL}\phi_{G}^{2} + \mathcal{O}({\phi_{G}^{3}})
\eea
where $f_{NL}$ is the parameter that characterises the local non-Gaussianity.

On the large scales relevant to this work, where the fluctuations are linear, the electron overdensity $\delta_e$ is related to the total matter overdensity field $\delta_m$ by a linear, scale-independent bias
\bea\label{electron overdensity equation}
\delta_e(\hat{n},z) = b_e(z)\,\delta_m(\hat{n},z)
\eea
where $b_e(z)$ is the electron bias. Similarly, the galaxy overdensity $\delta_g$ is related to the same matter overdensity field $\delta_m$ through a linear, scale-independent bias on large scales
\bea\label{galaxy overdensity equation}
\delta_g(\hat{n},z) = b_g(z)\,\delta_m(\hat{n},z)
\eea
where $b_g(z)$ is the linear galaxy bias.

The quadratic term in eqn \ref{non gaussianity field equation} couples the long-wavelength potential to the small-scale density fluctuations that govern the formation of collapsed structures. Hence, for a tracer $X$, the bias $b_X(z)$, gets a correction term
\bea\label{bias correction equation}
b_X(z) \;\longrightarrow\; b_X(z) + \Delta b_X(k,z)
\eea
where the non-Gaussian correction is \cite{DesjacquesJeongSchmidt, Dalal2008, MatarreseVerde2008, Slosar2008}
\bea\label{non gaussian bias correction equation}
\Delta b_X(k,z) = 3f_{NL}\,\delta_{c}\[b_X(z) - 1\]
\frac{\Omega_{m0}\,H_0^{2}}{c^{2}\,k^{2}\,T(k)\,D(z)}
\eea
where $\delta_c \simeq 1.686$ is the critical linear overdensity for spherical collapse, $k$ is the wavenumber, $T(k)$ is the matter transfer function normalised to unity as $k \to 0$, and $D(z)$ is the linear growth factor normalised to be $(1+z)^{-1}$ in the matter domination era.

One can see from eqn \ref{non gaussian bias correction equation} that the correction scales as $k^{-2}$; hence, it grows rapidly at large scales, and the $f_{NL}$ signal is concentrated at the lowest accessible wavenumbers. The amplitude of the correction is also proportional to $(b_X - 1)$. Hence, the effect is strongest for highly biased tracers and vanishes for an unbiased tracer with $b_X = 1$. Since galaxies are strongly biased $(b_g > 1)$, their non-Gaussian response is large and positive, whereas the free-electron field on large scales $(b_e < 1)$ responds weakly and with the opposite sign \cite{ShawRuddNagai, mccarty2026}. The power spectrum, $P_{AB}(k,z)$ of two projected fields, $A(\hat{n})$ and $B(\hat{n})$ is defined as
\bea\label{power spectrum general definition equation}
\langle \delta_{A}(k,z) \delta_{B}(k',z)\rangle = \( 2\pi \)^{3}\delta_{D}(k+k')P_{AB}(k,z)
\eea
where $\delta_{D}$ is the Dirac delta function, $\delta_{A}(k,z)$ and $\delta_{B}(k',z)$ are the respective Fourier transforms of the density contrast of fields $A$ and $B$. Using eqn \ref{power spectrum general definition equation}, the electron auto-power spectrum $P_{ee}(k,z)$, can be written as
\bea\label{electron power spectrum equation}
P_{ee}(k,z) = [b_{e} + \Delta b_{e}(k,z)]^{2} P_{mm}^{1}(k,z)
\eea
where $P_{mm}^{1}(k,z)$ is the linear matter power spectrum. Similarly, the galaxy auto-power spectrum $P_{gg}(k,z)$, can be written as
\bea\label{galaxy power spectrum equation}
P_{gg}(k,z) = [b_{g} + \Delta b_{g}(k,z)]^{2} P_{mm}^{1}(k,z)
\eea
and the electron-galaxy cross-power spectrum can be written as
\bea\label{electron galaxy power spectrum equation}
P_{eg}(k,z) = [b_{e} + \Delta b_{e}(k,z)][b_{g} + \Delta b_{g}(k,z)] P_{mm}^{1}(k,z)
\eea

\subsection{Angular power spectrum}\label{subsection Angular power spectrum}
%
The angular cross-power spectrum $C^{AB}(\ell)$ between two projected fields, $A(\hat{n})$ and $B(\hat{n})$ is defined as
\bea\label{general angular power spectrum equation}
\langle a_{\ell m}^{A} a_{\ell'm'}^{B\ast} \rangle = \delta^{K}_{\ell\ell'}\delta^{K}_{mm'}C^{AB}(\ell)
\eea
where $\delta^{K}_{ij}$ is the Kronecker delta function, and $^\ast$ denotes the complex conjugate. $a_{\ell m}^{A}$ and $a_{\ell'm'}^{B}$ are the spherical harmonic coefficients which are computed by expanding the density contrast fields in Fourier space followed by taking the spherical harmonic transform and expanding the exponential into plane waves to get
\bea\label{spherical harmonic coefficients equation}
a_{\ell,m}^{A} = 4\pi i^{\ell} \int^{\chi_{H}}_{0} d\chi \, W_{A}(\chi) \int^{\infty}_{0} \frac{d^3k}{(2\pi)^{3}} \delta_{A}(k,z) \, j_{\ell}(k\chi) \, Y_{\ell m}^{\ast}(\hat{k})
\eea
where $j_{\ell}$ is the spherical Bessel function of the first kind of order $\ell$, and $Y_{\ell m}$ are the spherical harmonics. The weighting function $W_{A}(\chi)$ encodes how each field projects the three-dimensional density along the line of sight, and differs fundamentally between the two tracers. For the galaxies it is simply
\bea\label{galaxy kernel}
W_{g}(\chi) = n_{g}(\chi)
\eea
where $n_g(\chi)$ is the normalised galaxy comoving source distribution. However, for the dispersion measure, it is given by eqn \ref{DM average weighting function eqn}. Substituting eqn \ref{spherical harmonic coefficients equation} into
eqn \ref{general angular power spectrum equation} and using the definition of the power spectrum in eqn \ref{power spectrum general definition equation}, the orthogonality of the spherical harmonics reduces the angular cross-spectrum to
\bea\label{exact Cl equation}
C^{AB}(\ell) = \frac{2}{\pi} \int^{\chi_{H}}_{0} \d\chi_{1}\; W_{A}(\chi_{1}) \int^{\chi_{H}}_{0} \d\chi_{2}\; W_{B}(\chi_{2}) \nonumber \\
\times \int^{\infty}_{0} k^{2}\, \d k\;
\sqrt{P_{AB}(k,\chi_{1})P_{AB}(k,\chi_{2})}\, j_{\ell}(k\chi_{1})\, j_{\ell}(k\chi_{2})
\eea
where $A,B \in \{\mathcal{D}, g\}$, with $P_{\mathcal{D}\mathcal{D}} \equiv P_{ee}$ and $P_{\mathcal{D}g} \equiv P_{eg}$ (section \ref{Primordial Non-Gaussianity and bias}), and the unequal-time correlator is computed as the geometric mean of the power spectra at the two times \cite{Reischke2021}.
For the tomographic analysis, each sample is divided into bins and the per-bin kernels $G^{i}_{A}(\chi)$ are restricted to the $i$-th bin, so that the observables are the angular auto- and cross-spectra $C^{A_i B_j}(\ell)$ between all bin pairs. The construction of the bins and kernels is detailed in \cref{subsection Tomography}.

\subsection{Covariance matrices}\label{subsection covariance matrices}
Using Wick's theorem for Gaussian  fields, the Gaussian covariance for two multi-probe spectra can be written as \cite{KrauseEifler2017, Sharma2025}
\bea\label{covariance equation}
\text{Cov}\[C^{AB}(\ell_i), C^{CD}(\ell_j)\] =
\frac{\delta^{K}_{ij}}{(2\ell_i + 1)\Delta\ell \, f_{\text{sky}}}
\[\hat{C}^{AC}(\ell_i)\hat{C}^{BD}(\ell_j) + \hat{C}^{AD}(\ell_i)\hat{C}^{BC}(\ell_j)\]
\eea
where $\Delta \ell$ is the width used for binning the power spectra, $f_{\text{sky}}$ is the fraction of the sky common to both surveys, and $\hat{C}^{AB}(\ell)$ is the observed power spectrum including the noise contribution $N^{A}$
\bea\label{noise contribution equation}
\hat{C}^{AB}(\ell) = C^{AB}(\ell) + \delta^{K}_{AB} N^{A}(\ell)
\eea
The galaxy auto-correlation carries the shot noise
\bea\label{galaxy shot noise equation}
N^{g}_{a} = \frac{1}{\bar{n}^{a}_{g}}
\eea
where $\bar{n}^{a}_{g}$ is the mean angular
number density of galaxies in bin $a$. The DM auto-correlation carries white noise induced by variation in the dispersion within the host:
\bea\label{DM shot noise equation}
N^{\mathcal{D}}_{i} = 4\pi f_{\text{sky}}\, n_{\text{tomo}}\,
\frac{\sigma^{2}_{\text{host}}(\bar{z}_{i})}{N_{\text{FRB}}}
\eea
where $N_{\text{FRB}}$ is the total number of bursts, $n_{\text{tomo}}$ the number of DM tomographic bins, $\bar{z}_i$ is the mean redshift of bin $i$. The noise is diagonal in bin space as different tomographic bins are uncorrelated in their noise properties. As the host contribution and the galaxy shot noise are statistically independent, the DM-galaxy cross-spectrum carries no noise term as well, as is also evident from eqn \ref{noise contribution equation}.
%
Throughout our work, we have used $\Delta \ell = 1$. Using eqn \ref{covariance equation}, the covariance matrices for $C^{\mathcal{D}\mathcal{D}}$, $C^{gg}$, and $C^{g\mathcal{D}}$ are, respectively
\bea\label{covariance matrices for dd equation}
\text{Cov}_{\mathcal{D}\mathcal{D}}[\ell_{i},\ell_{j}] &=& \frac{2\delta^{K}_{ij}\left( C^{\mathcal{D}\mathcal{D}}(\ell_i) + N^{\mathcal{D}} \right)^2}{(2\ell_i + 1) \, f_{\text{sky}}}
\\
\text{Cov}_{gg}[\ell_{i},\ell_{j}] &=& \frac{2\delta^{K}_{ij}\( C^{gg}(\ell_i) + N^{g} \)^2}{(2\ell_i + 1) \, f_{\text{sky}}}\label{covariance matrices for gg equation}
\\
\text{Cov}_{g\mathcal{D}}[\ell_{i},\ell_{j}] &=& \frac{\delta^{K}_{ij}}{(2\ell_{i} + 1) \, f_{\text{sky}}} \left[ (C^{g\mathcal{D}}(\ell_i))^2 + (C^{\mathcal{D}\mathcal{D}}(\ell_i) + N^{\mathcal{D}})(C^{gg}(\ell_i) + N^{g}) \right]\label{covariance matrices for gd equation}
\eea
The data vectors for the joint analysis of $\text{g}\mathcal{D}+\mathcal{D}\mathcal{D}$ and $\text{g}\mathcal{D}+gg$ are $(C^{g\mathcal{D}}(\ell),C^{\mathcal{D}\mathcal{D}}(\ell))^{T}$ and $(C^{g\mathcal{D}}(\ell),C^{gg}(\ell))^{T}$ respectively and the corresponding covariance matrices are given by
\bea\label{covariance matrices for gd+DD equation}
\text{Cov}_{g\mathcal{D}+\mathcal{D}\mathcal{D}} = \begin{pmatrix}
  \text{Cov}_{g\mathcal{D}}[\ell_{i},\ell_{j}] & \text{Cov}_{g\mathcal{D} - \mathcal{D}\mathcal{D}}[\ell_{i},\ell_{j}] \\
  \text{Cov}_{g\mathcal{D} - \mathcal{D}\mathcal{D}}[\ell_{i},\ell_{j}] & \text{Cov}_{\mathcal{D}\mathcal{D}}[\ell_{i},\ell_{j}]
\end{pmatrix}\\
\text{Cov}_{g\mathcal{D}+gg} = \begin{pmatrix}
  \text{Cov}_{g\mathcal{D}}[\ell_{i},\ell_{j}] & \text{Cov}_{g\mathcal{D} - gg}[\ell_{i},\ell_{j}] \\
  \text{Cov}_{g\mathcal{D} - gg}[\ell_{i},\ell_{j}] & \text{Cov}_{gg}[\ell_{i},\ell_{j}]
\end{pmatrix}\label{covariance matrices for gd+gg equation}
\eea
where the off-diagonal terms in eqn \ref{covariance matrices for gd+DD equation} and eqn \ref{covariance matrices for gd+gg equation} are given by
\bea\label{cross covariance gD-DD equation}
\text{Cov}_{g\mathcal{D} - \mathcal{D}\mathcal{D}}[\ell_{i},\ell_{j}] &=& \frac{2\delta^{K}_{ij} \, C^{g\mathcal{D}}(\ell_i) \left( C^{\mathcal{D}\mathcal{D}}(\ell_i) + N^{\mathcal{D}} \right)}{(2\ell_i + 1) \, f_{\text{sky}}}
\\
\text{Cov}_{g\mathcal{D} - gg}[\ell_{i},\ell_{j}] &=& \frac{2\delta^{K}_{ij} \, C^{g\mathcal{D}}(\ell_i) \left( C^{gg}(\ell_i) + N^{g} \right)}{(2\ell_i + 1) \, f_{\text{sky}}}\label{cross covariance gD-gg equation}
\eea

Similarly, the data vector for the full $3\times2$ point joint analysis $(gg + g\mathcal{D} + \mathcal{D}\mathcal{D})$ is $(C^{gg(\ell)}, C^{g\mathcal{D}}(\ell),C^{\mathcal{D}\mathcal{D}}(\ell))^{T}$ and the corresponding covariance matrix is given by
\bea\label{covariance matrices for gg+gd+DD equation}
\text{Cov}_{gg + g\mathcal{D}+\mathcal{D}\mathcal{D}} = \begin{pmatrix}
  \text{Cov}_{gg}[\ell_{i},\ell_{j}] & \text{Cov}_{gg - g\mathcal{D}}[\ell_{i},\ell_{j}] & \text{Cov}_{gg - \mathcal{D}\mathcal{D}}[\ell_{i},\ell_{j}] \\
  \text{Cov}_{gg - g\mathcal{D}}[\ell_{i},\ell_{j}] & \text{Cov}_{g\mathcal{D}}[\ell_{i},\ell_{j}] & \text{Cov}_{g\mathcal{D} - \mathcal{D}\mathcal{D}}[\ell_{i},\ell_{j}] \\
  \text{Cov}_{gg - \mathcal{D}\mathcal{D}}[\ell_{i},\ell_{j}] & \text{Cov}_{g\mathcal{D} - \mathcal{D}\mathcal{D}}[\ell_{i},\ell_{j}] & \text{Cov}_{\mathcal{D}\mathcal{D}}[\ell_{i},\ell_{j}]
\end{pmatrix}
\eea
where the cross variance terms in eqn \ref{covariance matrices for gg+gd+DD equation} are given by
\bea\label{cross covariance gg-gD equation}
\text{Cov}_{gg - g\mathcal{D}}[\ell_{i},\ell_{j}] &=& \frac{2\delta^{K}_{ij} \left( C^{gg}(\ell_i) + N^{g} \right) \, C^{g\mathcal{D}}(\ell_i)}{(2\ell_i + 1) \, f_{\text{sky}}}\\
\text{Cov}_{gg - \mathcal{D}\mathcal{D}}[\ell_{i},\ell_{j}] &=& \frac{2\delta^{K}_{ij} \(C^{g\mathcal{D}}(\ell_i)\)^2 }{(2\ell_i + 1) \, f_{\text{sky}}}\label{cross covariance gg-DD equation}
\eea

%

\subsection{Tomographic analysis}\label{subsection Tomography}
\begin{figure}[t]
\centering
\includegraphics[width=\textwidth]{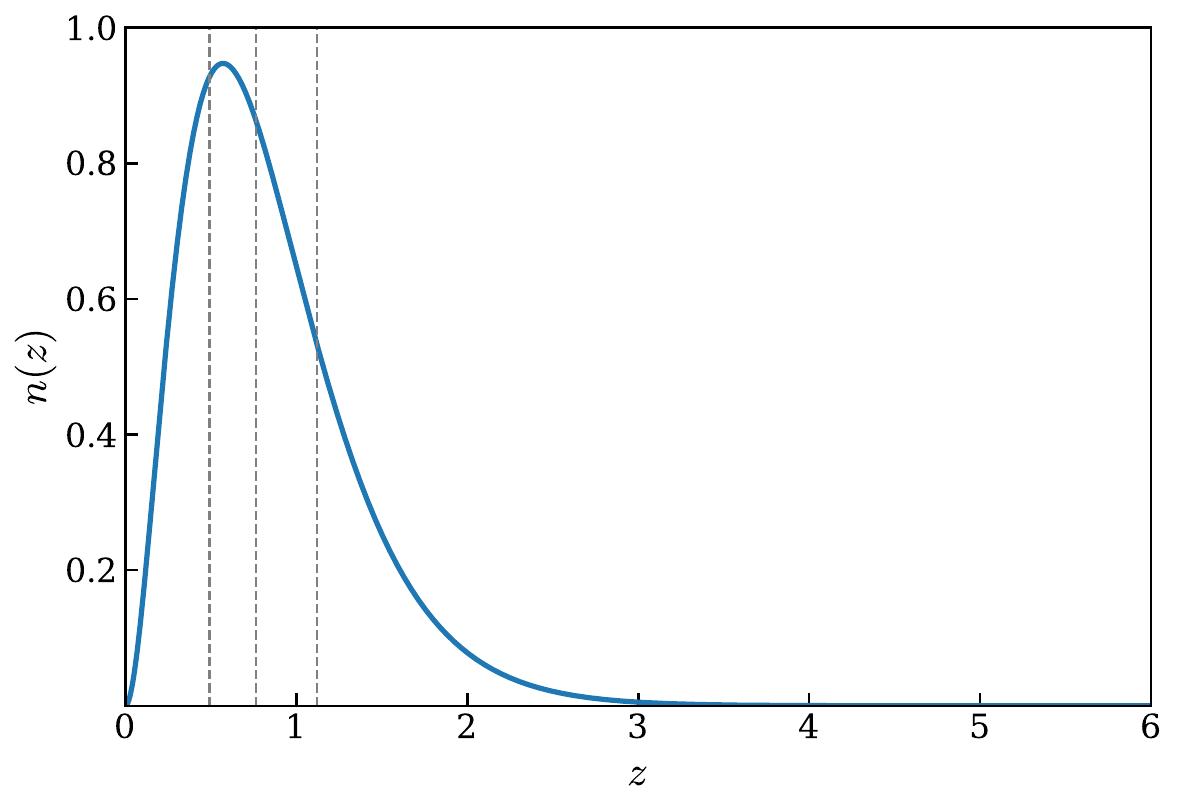}
\caption{The FRB source distribution $n(\text{z})$ as a function of redshift, with the four tomographic bins.}
\label{fig:nz}
\end{figure}

%
As the redshift distribution of FRBs is assumed to be broad, the DM weighting function $W_{\mathcal{D}}(\chi)$ spans a wide range of comoving distances. The line-of-sight projection smooths out the scale-dependent-bias imprint of eqn~\ref{non gaussian bias correction equation}, which is concentrated at low wavenumbers where the $f_{NL}$ signal resides. This makes the signal harder to isolate in the projected angular spectrum $C^{\mathcal{D}\mathcal{D}}(\ell)$. To improve the sensitivity, we divide the sample into $n_{\text{tomo}}$ tomographic bins. The sample is split at equal-source-count edges of the $n(z)$ distribution, as shown in Figure \ref{fig:nz}. The weighting function in the $i$-th bin is then obtained by restricting the source integral in eqn \ref{DM average weighting function eqn} to that bin,
\bea\label{per bin DM kernel equation}
G^{i}_{\mathcal{D}}(\chi) = W_{DM}(\chi)\,
\frac{\int_{\max(\chi,\chi_i)}^{\chi_{i+1}} n(\chi')\,\d\chi'}
{\int_{\chi_i}^{\chi_{i+1}} n(\chi')\,\d\chi'}
\eea
where $\chi_i$ is the comoving distance boundary of the $i$-th bin and $W_{DM}(\chi) = W_{DM}(z) \left|\frac{dz}{d\chi}\right|$. Similarly, the galaxy sample is divided into equal-number bins in redshift, with the per-bin kernel
\bea\label{per bin galaxy kernel equation}
G^{a}_{g}(\chi) = \frac{n^{a}_{g}(\chi)}{\int_{\chi_a}^{\chi_{a+1}} n_{g}(\chi')\,\d\chi'}
\eea
where $n^{a}_{g}(\chi)$ is the galaxy source distribution restricted to bin $a$. The auto- and cross-spectra $C^{X_i Y_j}(\ell)$ between all bin pairs then follow from eqn~\ref{exact Cl equation} with these per-bin kernels.

\subsection{Signal-to-noise ratio and Fisher forecasts}\label{subsection Fisher}
%
Collecting the tomographic signal spectra at a given multipole $\ell$ into the matrix $\mathbf{S}(\ell)$, with elements $C^{X_i Y_j}(\ell)$, and the shot noise into the matrix $\mathbf{N}$, such that $\mathbf{C}(\ell) = \mathbf{S}(\ell) + \mathbf{N}$, the per-multipole SNR is given by
\bea\label{snr per ell equation}
\text{SNR}^{2}(\ell) = \frac{2\ell+1}{2}\,f_{\text{sky}}\,
\text{Tr}\[\mathbf{S}(\ell)\,\mathbf{C}^{-1}(\ell)\,
\mathbf{S}(\ell)\,\mathbf{C}^{-1}(\ell)\]
\eea
and the cumulative signal-to-noise up to a multipole $\ell_{\max}$ is
\bea\label{snr cumulative equation}
\text{SNR}(\ell_{\max}) = \[\sum_{\ell \le \ell_{\max}} \text{SNR}^{2}(\ell)\]^{1/2}
\eea

\begin{figure}[t]
\centering
\includegraphics[width=\textwidth]{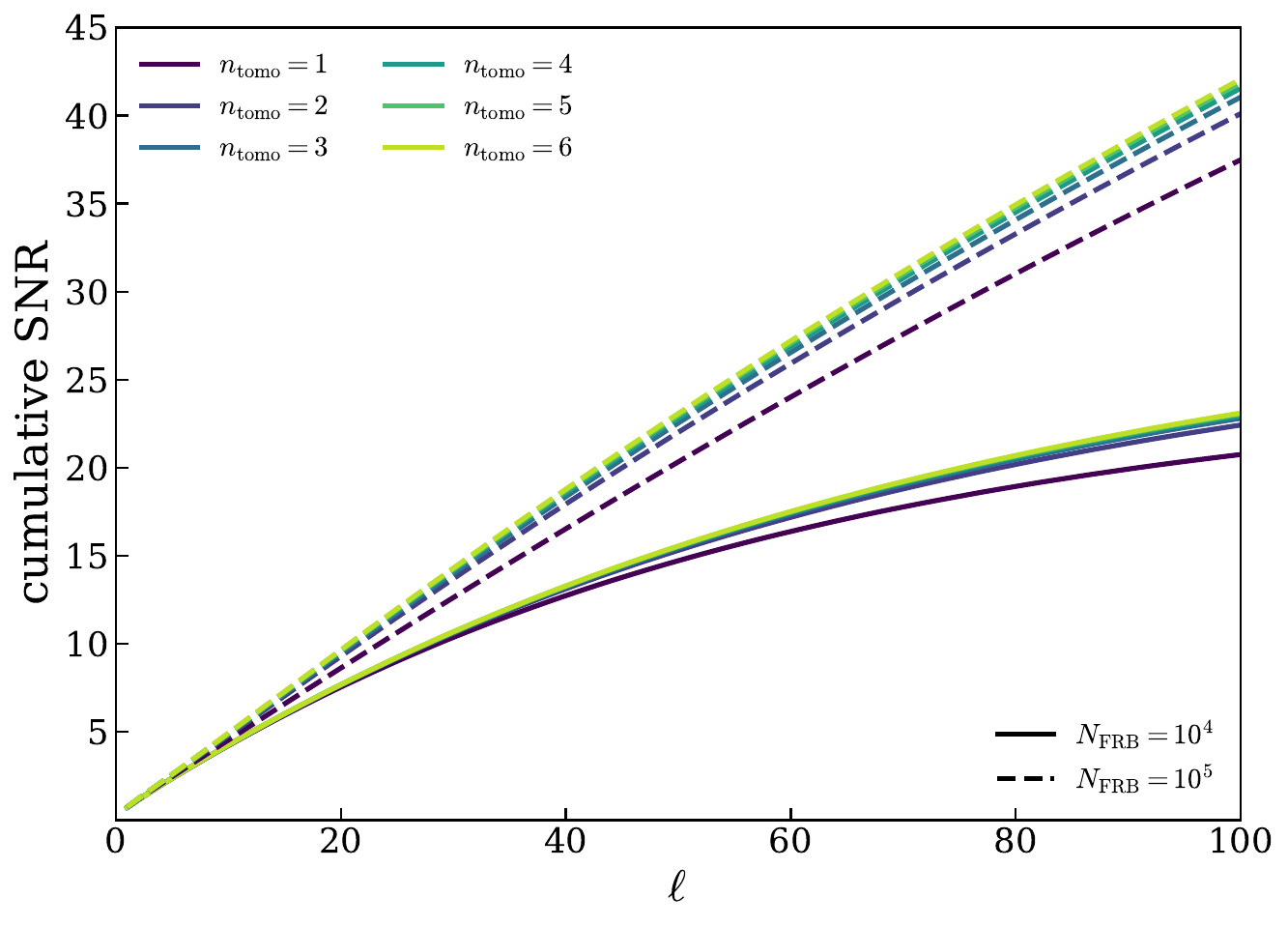}
\caption{Cumulative signal-to-noise ratio of the DM auto-spectrum as a function of $\ell_{\max}$, for different numbers of tomographic bins, for $N_{\text{FRB}} = 10^{4}$ (solid) and $10^{5}$ (dashed).}
\label{fig:snr}
\end{figure}

Figure~\ref{fig:snr} shows the cumulative SNR of the DM auto-spectrum for different numbers of tomographic bins. Adding bins increases the signal but also increases the correlation between bins, so the gain saturates. As Figure~\ref{fig:snr} shows, the improvement is negligible beyond $n_{\text{tomo}} \simeq 4$, which we adopt for all subsequent results.

To forecast the parameter uncertainties, we use the Fisher information matrix. Assuming a Gaussian likelihood, the Fisher matrix takes the form \cite{Tegmark1997, Reischke2021}
\bea\label{fisher trace equation}
F_{\alpha\beta} = f_{\text{sky}} \sum_{\ell = \ell_{\min}}^{\ell_{\max}}
\frac{2\ell+1}{2}\,
\text{Tr}\[\mathbf{C}^{-1}(\ell)\,\partial_{\alpha}\mathbf{C}(\ell)\,
\mathbf{C}^{-1}(\ell)\,\partial_{\beta}\mathbf{C}(\ell)\]\bigg|_{\theta = \theta_0}
\eea
where $\partial_{\alpha}$ denotes the derivative with respect to the parameter $\theta_{\alpha}$, evaluated at the fiducial point $\theta_0$. The noise $\mathbf{N}$ is independent of the parameters and hence, only the signal spectra are differentiated. Our baseline parameter set is $\theta = (f_{NL}, b_e^{0}, z_{\text{fb}}, b_g^{\text{amp}})$, comprising the primordial non-Gaussianity amplitude and the three bias parameters, with a fiducial $f_{NL} = 0$ and the cosmology held fixed at its \textit{Planck} 2018 values \cite{Planck2018params}. The fiducial values and models of the bias parameters are specified in \cref{section Survey specifications and numerical setup}. In \cref{subsection robustness}, we extend this set with the six $\Lambda$CDM parameters (using \textit{Planck} priors \cite{Planck2018params}) using the same Fisher matrix formalism to assess the impact of marginalising over cosmology.

The marginalised $1\sigma$ uncertainty on a parameter is obtained from the inverse Fisher matrix, $\sigma(\theta_{\alpha}) = \sqrt{(F^{-1})_{\alpha\alpha}}$, while the unmarginalised error, obtained by fixing the remaining parameters, is $1/\sqrt{F_{\alpha\alpha}}$. To disentangle the contribution of each observable to the $f_{NL}$ constraint, we evaluate the Fisher matrix for six separate combinations of the angular spectra: $\mathcal{D}\mathcal{D}$, $\mathcal{D}g$, $\mathcal{D}\mathcal{D} + \mathcal{D}g$, $\mathcal{D}g + gg$, $gg$, $\mathcal{D}\mathcal{D} + \mathcal{D}g + gg$. Each combination is built from its own data vector and covariance matrix, as described in \cref{subsection covariance matrices}.

\section{Survey specifications and numerical setup}\label{section Survey specifications and numerical setup}
%
We forecast the constraints on $f_{NL}$ for a joint analysis of an FRB survey and a galaxy survey with an overlapping sky fraction $f_{\text{sky}} = 0.35$. For the FRB sample, we adopt a source redshift distribution of the form $n(z) \propto z^{2}\exp(-\alpha z)$ with $\alpha = 3.5$ \cite{Reischke2021, Sharma2025}, and consider two representative sample sizes, $N_{\text{FRB}} = 10^{4}$ and $N_{\text{FRB}} = 10^{5}$, spanning the range from ongoing and next-generation wide-field radio surveys \cite{frbcollaboration2025chimefrboutriggersdesignoverview, CHORD, hallinan2019dsa2000radiosurvey, CARILLI2004979}. The dispersion measure is modelled to a maximum redshift $z_{\max} = 6$. The host scatter is taken as $\sigma_{\text{host}}(z) = 90\, (1+z)^{-1}\,\text{pc cm}^{-3}$ \cite{Sharma2025}. For $f_{\text{IGM}}(z)$, we adopt a slight redshift dependence, with $10\%$ ($20\%$) of baryons locked up in galaxies at $z \gtrsim 1.5$ ($z \lesssim 0.4$), and a linear interpolation in between \cite{ShullSmithDanforth, Reischke2021}.

For the galaxy survey, we adopt the specifications of the DESI DR1 LRG clustering sample, which includes 8 million objects over $36\%$ of the sky, spanning the redshift range $0.4 < z < 1.1$ with redshift distribution $n_g(z) \propto \bar{n}_g\,\d V/\d z$ and comoving number density $\bar{n}_g = 5\times10^{-4}\,h^{3}\,\text{Mpc}^{-3}$ \cite{Zhou}.

The galaxy bias is modelled as $b_g(z) = b_g^{\text{amp}}/D(z)$, and $b_g^{\text{amp}} = 1.7$ \cite{desicollaboration} . The electron bias is modelled as a linear ramp in redshift \cite{Reischke2021, dlamini2026fastradioburstdispersion}
\bea\label{electron bias ramp equation}
b_e(z) = b_e^{0} + \(1 - b_e^{0}\)\frac{z}{z_{\text{fb}}}
\eea
We adopt the fiducial values $b_e^{0} = 0.79$ and $z_{\text{fb}} = 5$ \cite{ShawRuddNagai, Reischke2021, dlamini2026fastradioburstdispersion}. We treat the three bias parameters $(b_g^{\text{amp}}, b_e^{0} ,z_{\text{fb}})$ as free and marginalise over them when constraining $f_{NL}$. Both samples are divided into $n_{\text{tomo}} = 4$ tomographic bins (\cref{subsection Fisher}).

We compute the linear matter power spectrum and transfer function using the Boltzmann code \textsc{class} \cite{CLASS} at a fixed fiducial cosmology based on the \textit{Planck} 2018 parameters \cite{Planck2018params}. When marginalising over cosmology, we recompute all cosmology-dependent quantities with \textsc{class} at each perturbed point, and impose independent Gaussian priors on the six $\Lambda$CDM parameters, with widths given by the \textit{Planck} 2018 marginalised uncertainties (TT, TE, EE+lowE+lensing) \cite{Planck2018params}.

\section{Results and discussions}\label{section Results and discussions}
%

\begin{figure}[h]
\centering
  \includegraphics[width=\textwidth]{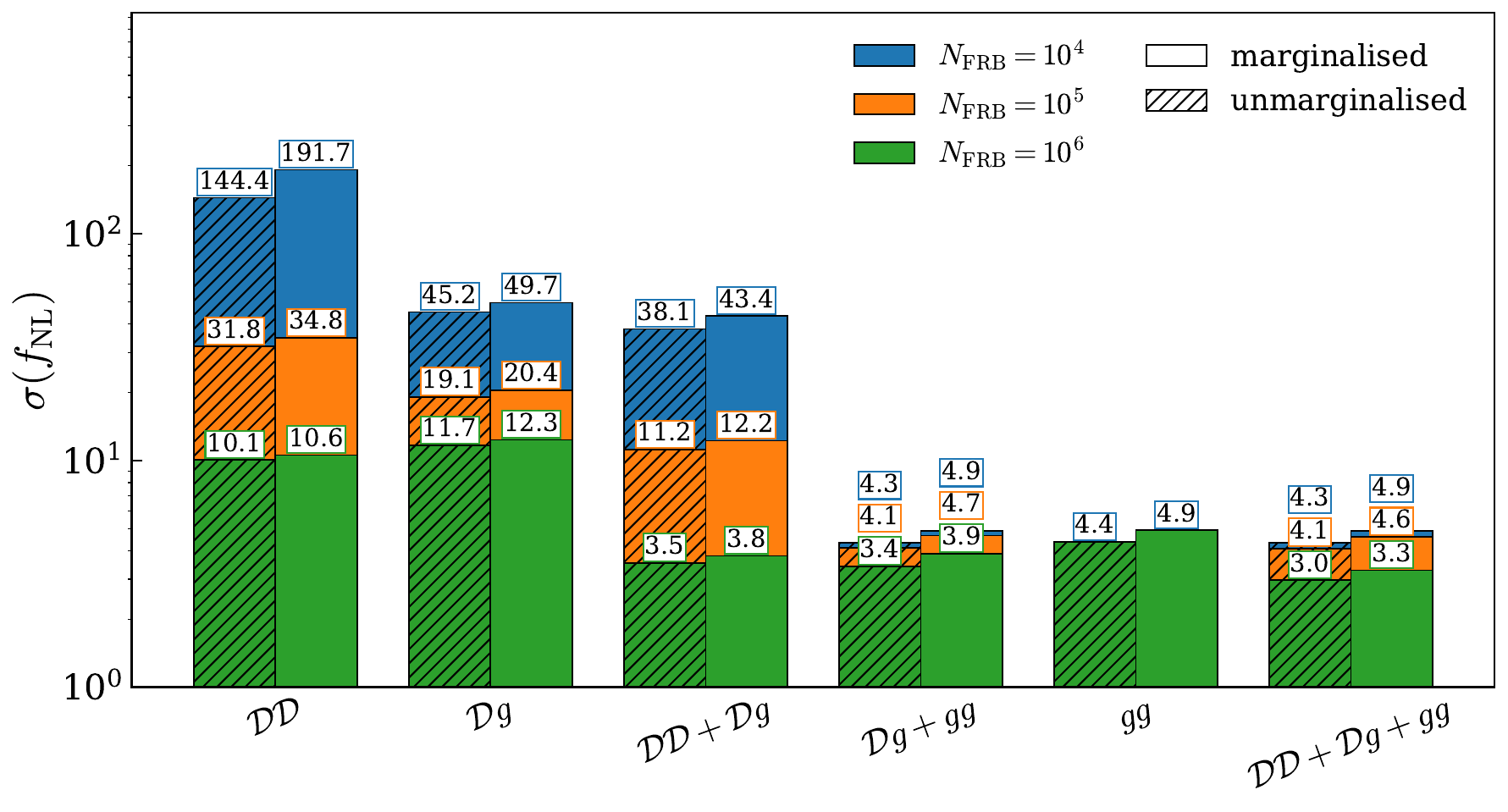}
\caption{$1\sigma$ constraints on $f_{NL}$ for the six observable combinations at fixed cosmology. Solid bars are marginalised over the bias parameters that each combination depends on whereas the hatched bars are unmarginalised. Bars for $N_{\text{FRB}} = 10^{4}$ (blue) and $10^{5}$ (orange) are overlaid with $N_{\text{FRB}} = 10^{6}$ (green) for comparison.}

\label{fig:barchart}
\end{figure}

We now present the forecast constraints on $f_{NL}$ from the six observable combinations, for the two FRB sample sizes $N_{\text{FRB}} = 10^{4}$ and $10^{5}$, at fixed cosmology and marginalised over the bias parameters relevant to each combination. Figure~\ref{fig:barchart} shows both the marginalised and unmarginalised constraints for the two sample sizes along with comparison at a larger burst count ($N_{\text{FRB}} = 10^{6}$).

\begin{figure}[h]
\centering
\begin{subfigure}{0.495\textwidth}
  \includegraphics[width=\textwidth]{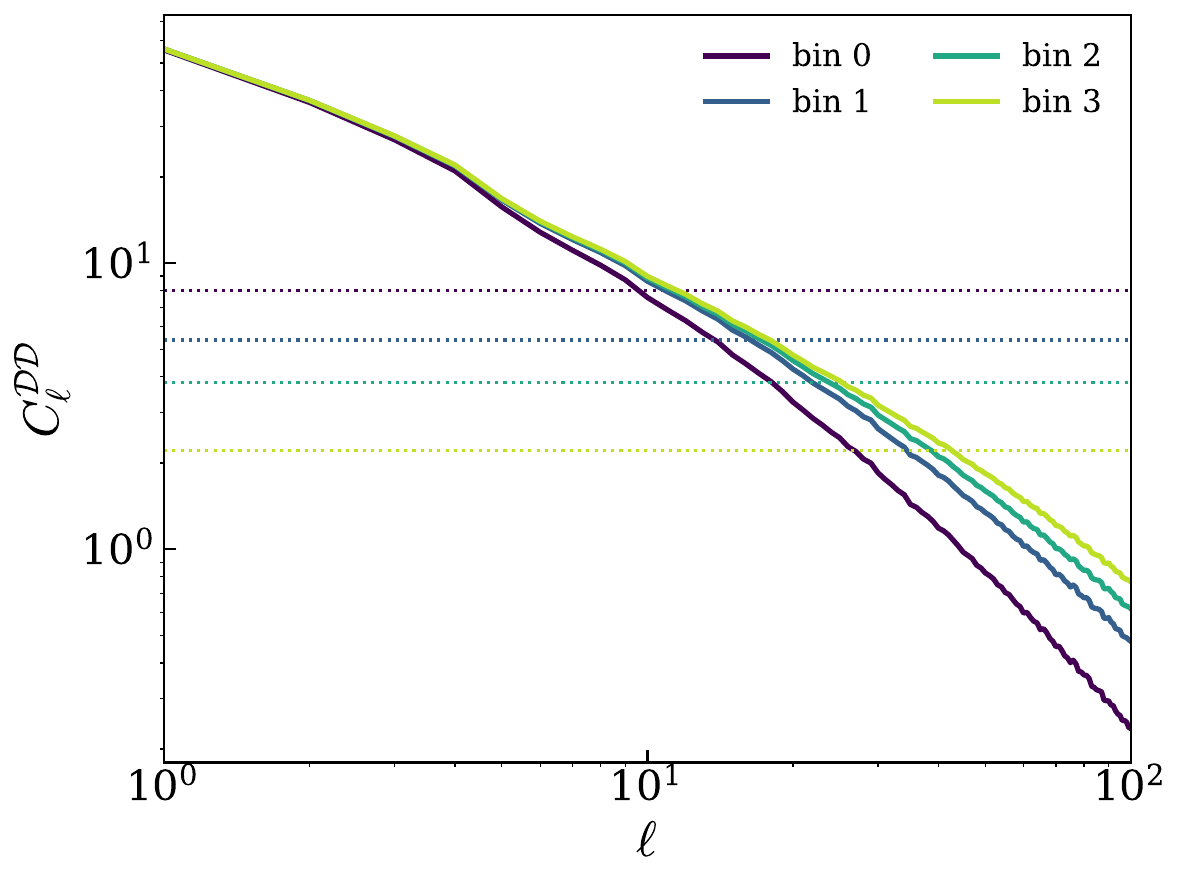}
  \caption{$N_{\text{FRB}} = 10^{4}$}
  \label{fig:cldd_noise_1e4}
\end{subfigure}
\begin{subfigure}{0.495\textwidth}
  \includegraphics[width=\textwidth]{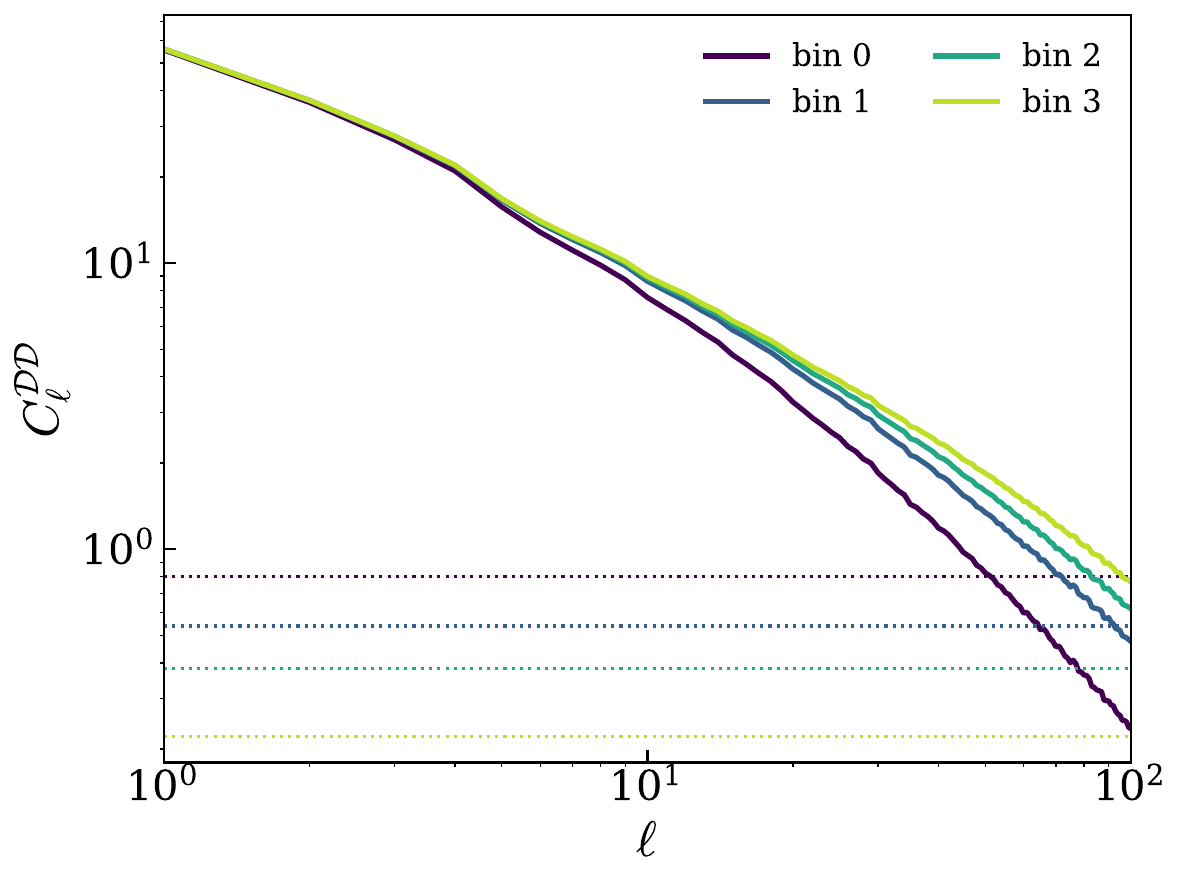}
  \caption{$N_{\text{FRB}} = 10^{5}$}
  \label{fig:cldd_noise_1e5}
\end{subfigure}
\caption{The DM auto-spectrum $C_\ell^{\mathcal{D}\mathcal{D}}$ (solid) and the host shot-noise floor $N_{i}^\mathcal{D}$ (dotted) in each tomographic bin, for $N_{\text{FRB}} = 10^{4}$ (left) and $10^{5}$ (right).}
\label{fig:cldd_noise}
\end{figure}

The results reveal a clear hierarchy. The DM auto-correlation is the weakest single probe, yielding $\sigma(f_{NL}) \simeq 192\,(35)$ for $N_{\text{FRB}} = 10^{4}\,(10^{5})$. This is because the correction to the electron bias scales as $(b_e - 1)$ (eqn \ref{non gaussian bias correction equation}), which is smaller than for galaxies. Additionally, at $10^{4}$ FRBs, the spectrum is dominated by noise above $\ell \sim 10$ in each tomographic bin (Figure \ref{fig:cldd_noise_1e4}). As the number of FRBs increases to  $10^{5}$, the noise in each tomographic bin decreases as $N^{\mathcal{D}}_{i} \propto N_{\text{FRB}}^{-1}$ (eqn \ref{DM shot noise equation}) and the signal starts to dominate over a wider range of multipoles (Figure \ref{fig:cldd_noise_1e5}), thus improving the constraint by nearly a factor of six.
The DM-galaxy cross-correlation improves on the DM auto-correlation, to $\sigma(f_{NL}) \simeq 50\,(20)$ for $N_{\text{FRB}} = 10^{4}\,(10^{5})$, but it remains well above the galaxy floor. The galaxy auto-correlation is by far the strongest single probe, $\sigma(f_{NL}) \simeq 5$, independent of $N_{\text{FRB}}$, owing to the large galaxy bias and the low shot noise of the dense, 8-million-object sample ($\bar{n}_g = 5\times10^{-4}\,h^{3}\,\text{Mpc}^{-3}$). Adding the DM auto to the cross ($\mathcal{D}\mathcal{D} + \mathcal{D}g$) gives $\sigma(f_{NL}) = 43 \, (12)$ for $N_{\text{FRB}} = 10^{4}\,(10^{5})$, which improves on either DM-only probe but remains above the galaxy floor as well at these two sample sizes. Appending the galaxy auto to the cross ($\mathcal{D}g + gg$) gives $\sigma(f_{NL}) \simeq 5$ for both $N_{\text{FRB}} = 10^{4}$ and $10^{5}$, which is the same as $gg$. The full $3\times2$-point analysis ($\mathcal{D}\mathcal{D} + \mathcal{D}g + gg$) attains $\sigma(f_{NL}) \simeq 5$, which lies essentially at the galaxy-only ($gg$) floor for both $N_{\text{FRB}} = 10^{4}$ and $10^{5}$. At these burst counts, the FRB sector adds little and its contribution with increasing burst counts is discussed in \cref{subsection nfrb dependence}.

\subsection{Dependence on the FRB sample size}\label{subsection nfrb dependence}

\begin{figure}[h]
\centering
\includegraphics[width=\textwidth]{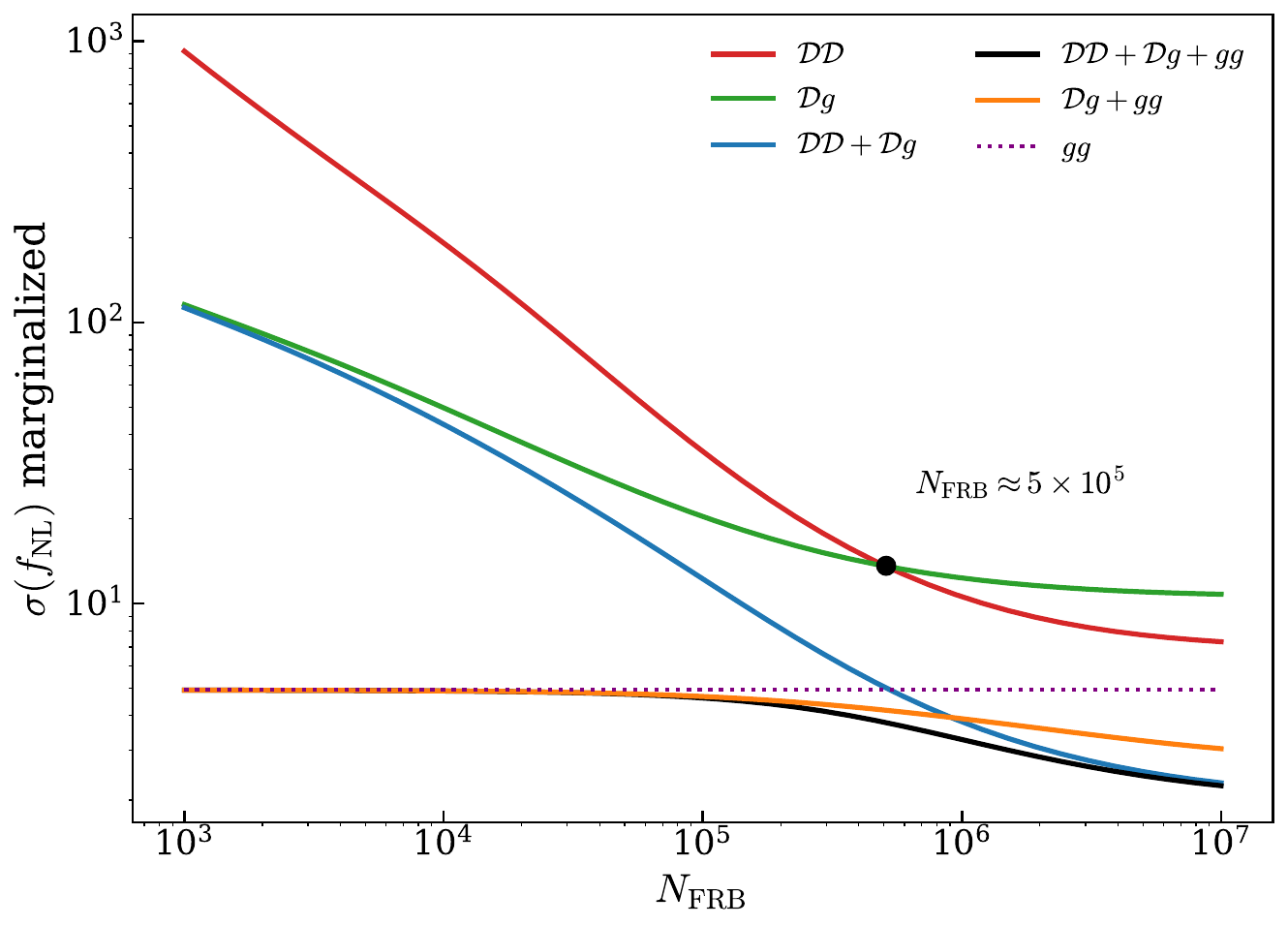}
\caption{Marginalised $\sigma(f_{NL})$ versus $N_{\text{FRB}}$ for the six observable combinations, at fixed cosmology, each marginalised over its relevant bias parameters.}\label{fig:mechanism}
\end{figure}

Figure~\ref{fig:mechanism} shows $\sigma(f_{NL})$ at fixed cosmology, marginalised over the bias parameters relevant to each combination, as a function of $N_{\text{FRB}}$ for all six cases: $\mathcal{D}\mathcal{D}, \, \mathcal{D}g , \, \mathcal{D}\mathcal{D}+\mathcal{D}g , \, \mathcal{D}\mathcal{D}+\mathcal{D}g+gg , \, \mathcal{D}g+gg \, \text{and} \, gg$. The DM auto-correlation and the DM-galaxy cross are limited by different terms in their covariance, so their $f_{NL}$ constraints scale differently with $N_{\text{FRB}}$. The $\mathcal{D}\mathcal{D}$ covariance is $\mathrm{Cov}_{\mathcal{D}\mathcal{D}} \propto (C^{\mathcal{D}\mathcal{D}} + N^{\mathcal{D}})^{2}$ (eqn~\ref{covariance matrices for dd equation}), with $N^{\mathcal{D}} \propto N_{\text{FRB}}^{-1}$ (eqn~\ref{DM shot noise equation}). Hence, at low $N_{\text{FRB}}$, the host noise dominates, and it decreases as the burst count grows. The constraint from the DM auto therefore falls steeply with burst count until $N^{\mathcal{D}}$ drops below $C^{\mathcal{D}\mathcal{D}}$, after which it approaches its own cosmic-variance floor. On the other hand, the host noise appears only linearly in the covariance for the DM-galaxy cross-spectrum ($\mathcal{D}g$) (eqn~\ref{covariance matrices for gd equation}),  $\mathrm{Cov}_{g\mathcal{D}} \propto (C^{g\mathcal{D}})^{2} + (C^{\mathcal{D}\mathcal{D}} + N^{\mathcal{D}})(C^{gg} + N^{g})$. The constraint from the cross thus improves more gently with $N_{\text{FRB}}$ and, once $N^{\mathcal{D}}$ becomes negligible, the covariance no longer depends on the burst count and the constraint saturates. The two curves cross at $N_{\text{FRB}} \approx 5\times10^{5}$. Below this value, the cross is the stronger FRB probe, and above it, the DM auto-correlation is stronger. The full $3\times2$-point constraint sits close to the galaxy floor up to $N_{\text{FRB}} \sim 10^{5}$ and departs from it at higher burst counts, reaching $\sigma(f_{NL}) \simeq 2$ at $N_{\text{FRB}} = 10^{7}$. The FRB sector is thus a subdominant addition to a galaxy-dominated measurement at $N_{\text{FRB}} \lesssim 10^{5}$, but becomes the decisive component once the DM shot noise falls below the signal. We note that $\mathcal{D}\mathcal{D}+\mathcal{D}g$ and the full $3\times2$-point continue to improve with $N_{\text{FRB}}$ and converge to one another, while the $\mathcal{D}g+gg$ combination, which lacks the DM auto, reaches only $\sigma(f_{NL}) \simeq 3$ at $N_{\text{FRB}} = 10^{7}$ compared to $\sigma(f_{NL}) \simeq 2$ from the full $3\times2$-point statistic at the same burst counts. This is because the DM-galaxy cross carries only the mode content shared between the two fields, so the DM-specific information accessible only through the DM auto-spectrum cannot be recovered from the cross and the galaxy auto alone.

\subsection{Joint constraints on bias and cosmology}\label{subsection robustness}


\begin{figure}[h]
\centering
\includegraphics[width=\textwidth]{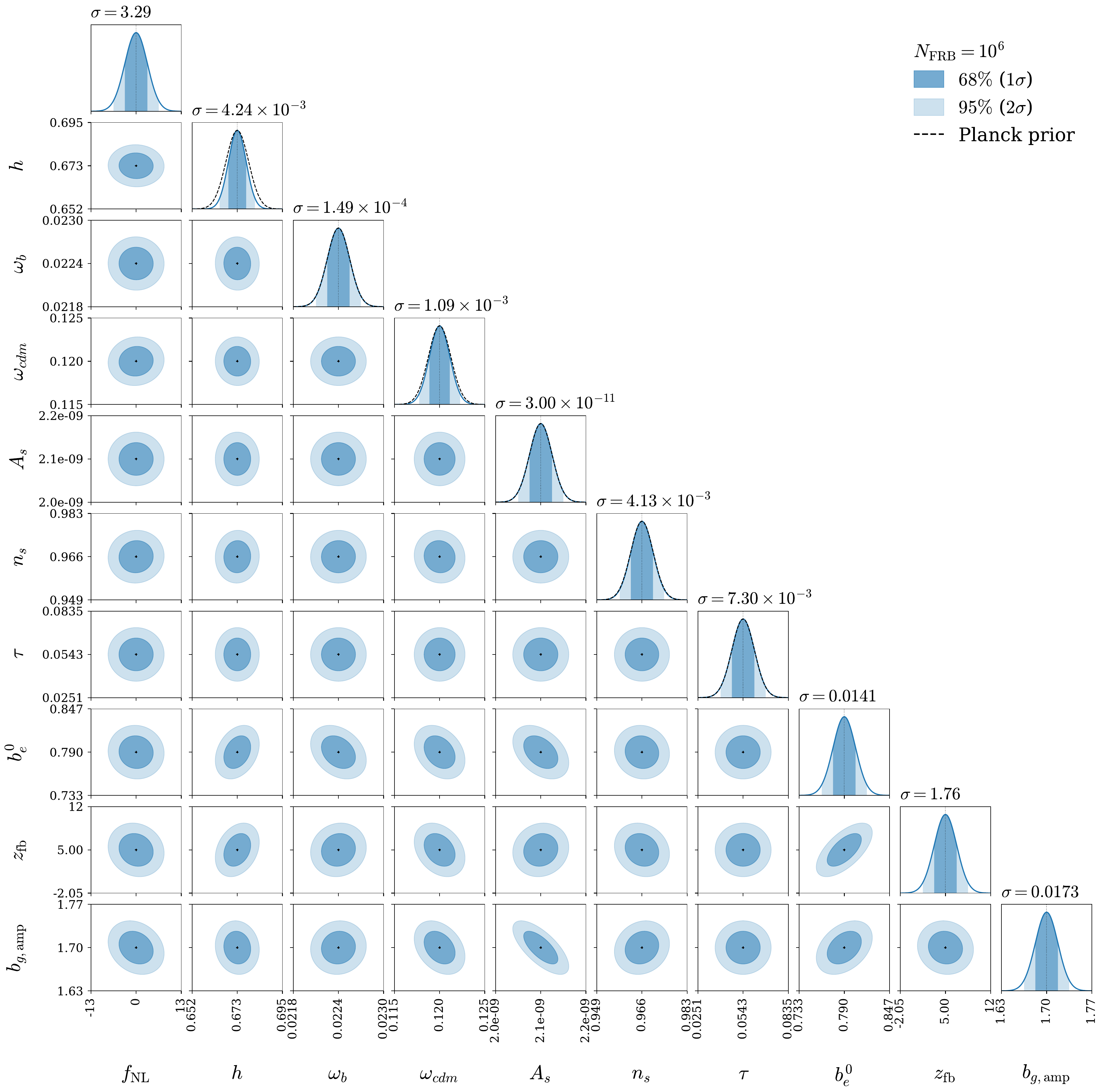}
\caption{Marginalised constraints on $f_{NL}$, the six $\Lambda$CDM parameters (with \textit{Planck} priors), and the three bias parameters from the full $3\times2$-point statistic, for $N_{\text{FRB}} = 10^{6}$.Off-diagonal panels show the $68\%$ and $95\%$ joint confidence regions. Diagonal panels show the one-dimensional marginalised distributions with the $1\sigma$ and $2\sigma$ intervals shaded and the marginalised $1\sigma$ error quoted above each panel, and the dashed curves show the \textit{Planck} priors.}
\label{fig:corner}
\end{figure}

To assess the impact of cosmological uncertainty, and to quantify what the full analysis constrains beyond $f_{NL}$, we extend the same $3\times2$-point statistic to a joint ten-parameter Fisher matrix comprising $f_{NL}$, the six base-$\Lambda$CDM parameters $(\omega_b, \omega_{\text{cdm}}, h, A_s, n_s, \tau)$ with \textit{Planck} priors \cite{Planck2018params}, and the three bias parameters $(b_e^{0}, z_{\text{fb}}, b_g^{\text{amp}})$. The resulting marginalised constraints for $N_{\text{FRB}} = 10^{6}$ are shown in Figure ~\ref{fig:corner}. We find that $\sigma(f_{NL})$ remains almost unchanged from its fixed-cosmology value (Figure~\ref{fig:barchart}), and hence the constraint on $f_{NL}$ is robust to this marginalisation. 

Beyond $f_{NL}$, the full $3\times2$-point analysis also constrains the bias parameters. From Figures~\ref{fig:corner}, we find $\sigma(b_g^{\text{amp}}) \simeq 0.017$, $\sigma(b_e^{0}) \simeq 0.0014$, and $\sigma(z_{\text{fb}}) \simeq 1.8$ for $N_{\text{FRB}} = 10^{6}$. Hence, the galaxy bias amplitude is measured to about one per cent and the present-day electron bias to about two per cent. Of the six $\Lambda$CDM parameters, $h$ is the most improved beyond its \textit{Planck} prior (by about $20\%$). The remaining cosmological parameters are essentially unconstrained by these data and stay almost at their priors. As seen in Figures~\ref{fig:corner}, $A_s$ is degenerate with the galaxy bias amplitude $b_g^{\rm amp}$, visible as the tilted $A_s$ -- $b_g^{\rm amp}$ contour.

\section{Conclusions}\label{section Conclusions}
%
We have forecast the sensitivity of the FRB-galaxy $3\times2$-point statistic to local primordial non-Gaussianity, combining the angular DM auto-spectrum, the DM-galaxy cross-spectrum, and the galaxy auto-spectrum for a DESI-like LRG sample \cite{Zhou} and an FRB population over $35\%$ of the sky. The full $3\times2$-point analysis constrains $\sigma(f_{NL}) \simeq 5$ for both $N_{\text{FRB}} = 10^{4}$ and $10^{5}$, marginalised over the three bias parameters at fixed cosmology (Figure~\ref{fig:barchart}). The DM auto and the DM-galaxy cross are limited by distinct terms in their covariance (\cref{subsection covariance matrices}), so their $f_{NL}$ constraints scale differently with burst count and the two cross at $N_{\text{FRB}} \approx 5\times10^{5}$ (Figure~\ref{fig:mechanism}). The joint constraint sits essentially at the galaxy floor for $N_{\text{FRB}} \lesssim 10^{5}$ but departs from it at higher burst counts and reaches $\sigma(f_{NL}) \simeq 3.3$ at $N_{\text{FRB}} = 10^{6}$ and $\sigma(f_{NL}) \simeq 2$ at $N_{\text{FRB}} = 10^{7}$, attainable in the SKA era \cite{wei2024SKA}.

Beyond $f_{NL}$, the full analysis simultaneously constrains the astrophysical bias parameters (\cref{subsection robustness}); at $N_{\text{FRB}} = 10^{6}$, the galaxy bias amplitude to about one per cent and the present-day electron bias to about two per cent. The analysis also improves $h$ over its \textit{Planck} prior by about $20\%$. The $f_{NL}$ constraint itself is robust to marginalisation over cosmology. The FRB-galaxy $3\times2$-point statistic thus strengthens $f_{NL}$ constraints beyond the galaxy sector alone, and its constraining power will grow with the burst counts of forthcoming wide-field FRB surveys.

\section{Acknowledgment}\label{section Acknowledgment}
We thank Brand Fortner for support. This work made use of the Illinois Campus Cluster, which is operated by the Illinois Campus Cluster Program (ICCP) and the National Center for Supercomputing Applications (NCSA) and which is supported by funds
from the University of Illinois at Urbana-Champaign.

\bibliographystyle{JHEP}
\bibliography{biblio}

\providecommand{\href}[2]{#2}\begingroup\raggedright\begin{thebibliography}{10}

\bibitem{Maldacena2003}
J.~Maldacena, \emph{Non-gaussian features of primordial fluctuations in single field inflationary models}, \href{https://doi.org/10.1088/1126-6708/2003/05/013}{\emph{Journal of High Energy Physics} {\bfseries 2003} (2003) 013–013}.

\bibitem{Bartolo2004}
N.~Bartolo, E.~Komatsu, S.~Matarrese and A.~Riotto, \emph{Non-gaussianity from inflation: theory and observations}, \href{https://doi.org/10.1016/j.physrep.2004.08.022}{\emph{Physics Reports} {\bfseries 402} (2004) 103–266}.

\bibitem{Planck2018NG}
{\scshape Planck} collaboration, \emph{{Planck 2018 results. IX. Constraints on primordial non-Gaussianity}}, \href{https://doi.org/10.1051/0004-6361/201935891}{\emph{Astron. Astrophys.} {\bfseries 641} (2020) A9} [\href{https://arxiv.org/abs/1905.05697}{{\ttfamily 1905.05697}}].

\bibitem{Dalal2008}
N.~Dalal, O.~Dore, D.~Huterer and A.~Shirokov, \emph{{The imprints of primordial non-gaussianities on large-scale structure: scale dependent bias and abundance of virialized objects}}, \href{https://doi.org/10.1103/PhysRevD.77.123514}{\emph{Phys. Rev. D} {\bfseries 77} (2008) 123514} [\href{https://arxiv.org/abs/0710.4560}{{\ttfamily 0710.4560}}].

\bibitem{MatarreseVerde2008}
S.~Matarrese and L.~Verde, \emph{{The effect of primordial non-Gaussianity on halo bias}}, \href{https://doi.org/10.1086/587840}{\emph{Astrophys. J. Lett.} {\bfseries 677} (2008) L77} [\href{https://arxiv.org/abs/0801.4826}{{\ttfamily 0801.4826}}].

\bibitem{Slosar2008}
A.~Slosar, C.~Hirata, U.~Seljak, S.~Ho and N.~Padmanabhan, \emph{{Constraints on local primordial non-Gaussianity from large scale structure}}, \href{https://doi.org/10.1088/1475-7516/2008/08/031}{\emph{JCAP} {\bfseries 08} (2008) 031} [\href{https://arxiv.org/abs/0805.3580}{{\ttfamily 0805.3580}}].

\bibitem{Mueller2022}
{\scshape eBOSS} collaboration, \emph{{Primordial non-Gaussianity from the completed SDSS-IV extended Baryon Oscillation Spectroscopic Survey II: measurements in Fourier space with optimal weights}}, \href{https://doi.org/10.1093/mnras/stac812}{\emph{Mon. Not. Roy. Astron. Soc.} {\bfseries 514} (2022) 3396} [\href{https://arxiv.org/abs/2106.13725}{{\ttfamily 2106.13725}}].

\bibitem{Chaussidon2024}
E.~Chaussidon et~al., \emph{{Constraining primordial non-Gaussianity with DESI 2024 LRG and QSO samples}}, \href{https://doi.org/10.1088/1475-7516/2025/06/029}{\emph{JCAP} {\bfseries 06} (2025) 029} [\href{https://arxiv.org/abs/2411.17623}{{\ttfamily 2411.17623}}].

\bibitem{Sullivan2023}
J.M.~Sullivan, T.~Prijon and U.~Seljak, \emph{Learning to concentrate: multi-tracer forecasts on local primordial non-gaussianity with machine-learned bias}, \href{https://doi.org/10.1088/1475-7516/2023/08/004}{\emph{Journal of Cosmology and Astroparticle Physics} {\bfseries 2023} (2023) 004}.

\bibitem{Lorimer2007}
D.R.~Lorimer, M.~Bailes, M.A.~McLaughlin, D.J.~Narkevic and F.~Crawford, \emph{{A bright millisecond radio burst of extragalactic origin}}, \href{https://doi.org/10.1126/science.1147532}{\emph{Science} {\bfseries 318} (2007) 777} [\href{https://arxiv.org/abs/0709.4301}{{\ttfamily 0709.4301}}].

\bibitem{Petroff2022}
E.~Petroff, J.W.T.~Hessels and D.R.~Lorimer, \emph{{Fast radio bursts at the dawn of the 2020s}}, \href{https://doi.org/10.1007/s00159-022-00139-w}{\emph{Astron. Astrophys. Rev.} {\bfseries 30} (2022) 2} [\href{https://arxiv.org/abs/2107.10113}{{\ttfamily 2107.10113}}].

\bibitem{Macquart2020}
J.P.~Macquart et~al., \emph{{A census of baryons in the Universe from localized fast radio bursts}}, \href{https://doi.org/10.1038/s41586-020-2300-2}{\emph{Nature} {\bfseries 581} (2020) 391} [\href{https://arxiv.org/abs/2005.13161}{{\ttfamily 2005.13161}}].

\bibitem{McQuinnAJL}
M.~McQuinn, \emph{{Locating the ''missing'' baryons with extragalactic dispersion measure estimates}}, \href{https://doi.org/10.1088/2041-8205/780/2/L33}{\emph{Astrophys. J. Lett.} {\bfseries 780} (2014) L33} [\href{https://arxiv.org/abs/1309.4451}{{\ttfamily 1309.4451}}].

\bibitem{Sharma2025}
K.~Sharma, E.~Krause, V.~Ravi, R.~Reischke, L.~Connor, P.R.~S. et~al., \emph{{Probing Baryonic Feedback and Cosmology with the 3 {\texttimes} 2-point Statistic of Fast Radio Bursts and Galaxies}}, \href{https://doi.org/10.3847/1538-4357/ae2ff9}{\emph{Astrophys. J.} {\bfseries 998} (2026) 109} [\href{https://arxiv.org/abs/2509.05866}{{\ttfamily 2509.05866}}].

\bibitem{Reischke2021}
R.~Reischke, S.~Hagstotz and R.~Lilow, \emph{{Probing primordial non-Gaussianity with Fast Radio Bursts}}, \href{https://doi.org/10.1103/PhysRevD.103.023517}{\emph{Phys. Rev. D} {\bfseries 103} (2021) 023517} [\href{https://arxiv.org/abs/2007.04054}{{\ttfamily 2007.04054}}].

\bibitem{Wang2025}
H.~Wang et~al., \emph{{Measurement of the Dispersion-Galaxy Cross-Power Spectrum with the Second CHIME/FRB Catalog}}, \href{https://doi.org/10.1103/9th9-qc51}{\emph{Phys. Rev. Lett.} {\bfseries 137} (2026) 041001} [\href{https://arxiv.org/abs/2506.08932}{{\ttfamily 2506.08932}}].

\bibitem{frbcollaboration2025chimefrboutriggersdesignoverview}
{\scshape CHIME/FRB} collaboration, \emph{{CHIME/FRB Outriggers: Design Overview}}, \href{https://doi.org/10.3847/1538-4357/adfdcc}{\emph{Astrophys. J.} {\bfseries 993} (2025) 55} [\href{https://arxiv.org/abs/2504.05192}{{\ttfamily 2504.05192}}].

\bibitem{CHORD}
K.~{Vanderlinde}, A.~{Liu}, B.~{Gaensler}, D.~{Bond}, G.~{Hinshaw}, C.~{Ng} et~al., \emph{{The Canadian Hydrogen Observatory and Radio-transient Detector (CHORD)}},  in \emph{Canadian Long Range Plan for Astronomy and Astrophysics White Papers}, vol.~2020, p.~28, Oct., 2019, \href{https://doi.org/10.5281/zenodo.3765414}{DOI} [\href{https://arxiv.org/abs/1911.01777}{{\ttfamily 1911.01777}}].

\bibitem{hallinan2019dsa2000radiosurvey}
G.~{Hallinan}, V.~{Ravi}, S.~{Weinreb}, J.~{Kocz}, Y.~{Huang}, D.P.~{Woody} et~al., \emph{{The DSA-2000 {\textemdash} A Radio Survey Camera}},  in \emph{Bulletin of the American Astronomical Society}, vol.~51, p.~255, Sept., 2019, \href{https://doi.org/10.48550/arXiv.1907.07648}{DOI} [\href{https://arxiv.org/abs/1907.07648}{{\ttfamily 1907.07648}}].

\bibitem{wei2024SKA}
J.-J.~Wei and C.-Y.~Gao, \emph{{Forecasts for Helium Reionization Detection with Fast Radio Bursts in the Era of Square Kilometre Array}},  \href{https://arxiv.org/abs/2409.01543}{{\ttfamily 2409.01543}}.

\bibitem{Planck2018params}
{\scshape Planck} collaboration, \emph{{Planck 2018 results. VI. Cosmological parameters}}, \href{https://doi.org/10.1051/0004-6361/201833910}{\emph{Astron. Astrophys.} {\bfseries 641} (2020) A6} [\href{https://arxiv.org/abs/1807.06209}{{\ttfamily 1807.06209}}].

\bibitem{Kulkarni2020}
S.R.~Kulkarni, \emph{{Dispersion measure: Confusion, Constants {\&} Clarity}},  \href{https://arxiv.org/abs/2007.02886}{{\ttfamily 2007.02886}}.

\bibitem{NE2001a}
J.M.~Cordes and T.J.W.~Lazio, \emph{{NE2001. 1. A New model for the galactic distribution of free electrons and its fluctuations}},  \href{https://arxiv.org/abs/astro-ph/0207156}{{\ttfamily astro-ph/0207156}}.

\bibitem{NE2001b}
J.M.~Cordes and T.J.W.~Lazio, \emph{{NE2001. 2. Using radio propagation data to construct a model for the galactic distribution of free electrons}},  \href{https://arxiv.org/abs/astro-ph/0301598}{{\ttfamily astro-ph/0301598}}.

\bibitem{ProchaskaZheng2019}
J.X.~Prochaska and Y.~Zheng, \emph{Probing galactic haloes with fast radio bursts}, \href{https://doi.org/10.1093/mnras/stz261}{\emph{Monthly Notices of the Royal Astronomical Society} {\bfseries 485} (2019) 648}.

\bibitem{DSA110CGM2023}
{\scshape Deep Synoptic Array Team} collaboration, \emph{{Deep Synoptic Array Science: A 50 Mpc Fast Radio Burst Constrains the Mass of the Milky Way Circumgalactic Medium}}, \href{https://doi.org/10.3847/1538-3881/adc725}{\emph{Astron. J.} {\bfseries 169} (2025) 330} [\href{https://arxiv.org/abs/2301.01000}{{\ttfamily 2301.01000}}].

\bibitem{Alonso2021}
D.~Alonso, \emph{{Linear anisotropies in dispersion-measure-based cosmological observables}}, \href{https://doi.org/10.1103/PhysRevD.103.123544}{\emph{Phys. Rev. D} {\bfseries 103} (2021) 123544} [\href{https://arxiv.org/abs/2103.14016}{{\ttfamily 2103.14016}}].

\bibitem{Wayland2026}
A.~Wayland, D.~Alonso and R.~Reischke, \emph{{Probing baryonic feedback with fast radio bursts: joint analyses with cosmic shear and galaxy clustering}}, \href{https://doi.org/10.1093/mnras/stag557}{\emph{Mon. Not. Roy. Astron. Soc.} {\bfseries 547} (2026) stag557} [\href{https://arxiv.org/abs/2602.12174}{{\ttfamily 2602.12174}}].

\bibitem{KomatsuSpergel}
E.~Komatsu and D.N.~Spergel, \emph{Acoustic signatures in the primary microwave background bispectrum}, \href{https://doi.org/10.1103/PhysRevD.63.063002}{\emph{Phys. Rev. D} {\bfseries 63} (2001) 063002}.

\bibitem{DesjacquesJeongSchmidt}
V.~Desjacques, D.~Jeong and F.~Schmidt, \emph{Large-scale galaxy bias}, \href{https://doi.org/https://doi.org/10.1016/j.physrep.2017.12.002}{\emph{Physics Reports} {\bfseries 733} (2018) 1}.

\bibitem{ShawRuddNagai}
L.D.~Shaw, D.H.~Rudd and D.~Nagai, \emph{Deconstructing the kinetic sz power spectrum}, \href{https://doi.org/10.1088/0004-637x/756/1/15}{\emph{The Astrophysical Journal} {\bfseries 756} (2012) 15}.

\bibitem{mccarty2026}
S.~McCarty et~al., \emph{{A new measurement of the FRB DM-galaxy cross correlation and a first joint analysis with the kinematic SZ effect}},  \href{https://arxiv.org/abs/2608.11296}{{\ttfamily 2608.11296}}.

\bibitem{KrauseEifler2017}
E.~Krause and T.~Eifler, \emph{{COSMOLIKE – cosmological likelihood analyses for photometric galaxy surveys}}, \href{https://doi.org/10.1093/mnras/stx1261}{\emph{Monthly Notices of the Royal Astronomical Society} {\bfseries 470} (2017) 2100}.

\bibitem{Tegmark1997}
M.~Tegmark, A.~Taylor and A.~Heavens, \emph{{Karhunen-Loeve eigenvalue problems in cosmology: How should we tackle large data sets?}}, \href{https://doi.org/10.1086/303939}{\emph{Astrophys. J.} {\bfseries 480} (1997) 22} [\href{https://arxiv.org/abs/astro-ph/9603021}{{\ttfamily astro-ph/9603021}}].

\bibitem{CARILLI2004979}
C.~Carilli and S.~Rawlings, \emph{Motivation, key science projects, standards and assumptions}, \href{https://doi.org/https://doi.org/10.1016/j.newar.2004.09.001}{\emph{New Astronomy Reviews} {\bfseries 48} (2004) 979}.

\bibitem{ShullSmithDanforth}
J.M.~{Shull}, B.D.~{Smith} and C.W.~{Danforth}, \emph{{The Baryon Census in a Multiphase Intergalactic Medium: 30\% of the Baryons May Still be Missing}}, \href{https://doi.org/10.1088/0004-637X/759/1/23}{\emph{The Astrophysical Journal} {\bfseries 759} (2012) 23} [\href{https://arxiv.org/abs/1112.2706}{{\ttfamily 1112.2706}}].

\bibitem{Zhou}
{\scshape DESI} collaboration, \emph{{Target Selection and Validation of DESI Luminous Red Galaxies}}, \href{https://doi.org/10.3847/1538-3881/aca5fb}{\emph{Astron. J.} {\bfseries 165} (2023) 58} [\href{https://arxiv.org/abs/2208.08515}{{\ttfamily 2208.08515}}].

\bibitem{desicollaboration}
{\scshape DESI} collaboration, \emph{{The DESI Experiment Part I: Science,Targeting, and Survey Design}},  \href{https://arxiv.org/abs/1611.00036}{{\ttfamily 1611.00036}}.

\bibitem{dlamini2026fastradioburstdispersion}
S.~Dlamini, \emph{{Fast radio burst dispersion measure-timing cross-correlations: bias self-calibration and primordial non-Gaussianity constraints}}, \href{https://doi.org/10.1140/epjc/s10052-026-16017-z}{\emph{Eur. Phys. J. C} {\bfseries 86} (2026) 779} [\href{https://arxiv.org/abs/2604.04897}{{\ttfamily 2604.04897}}].

\bibitem{CLASS}
D.~Blas, J.~Lesgourgues and T.~Tram, \emph{{The Cosmic Linear Anisotropy Solving System (CLASS). Part II: Approximation schemes}}, \href{https://doi.org/10.1088/1475-7516/2011/07/034}{\emph{Journal of Cosmology and Astroparticle Physics} {\bfseries 2011} (2011) 034–034}.

\end{thebibliography}\endgroup

\end{document}